\documentclass[pre,aps,floatfix,twocolumn,10pt]{revtex4}
\usepackage{graphicx}
\usepackage{subfigure}
\usepackage{amssymb}
\usepackage{amsmath}
\usepackage[dvips]{epsfig}
\usepackage[mathcal]{eucal}

\def\f{\phi}
\def\th{\theta}
\def\s{\sigma}
\def\0{{(0)}}
\def\1{{(1)}}
\def\csch{\textrm{csch}}
\def\a{\alpha}
\def\b{\beta}

\def\d{\partial}
\def\g{\gamma}

\begin{document}

\title{Elastic Instability Triggered Pattern Formation}

\author{Elisabetta A. Matsumoto}
\affiliation{Department of Physics and Astronomy, University of Pennsylvania, Philadelphia, PA 19104-6396, USA}
\author{Randall D. Kamien}
\affiliation{Department of Physics and Astronomy, University of Pennsylvania, Philadelphia, PA 19104-6396, USA}

\date{\today}

\begin{abstract}
Recent experiments have exploited elastic instabilities in membranes to create complex patterns.  However, the rational design of such structures poses many challenges, as they are products of nonlinear elastic behavior.  We pose a simple model for determining the orientational order of such patterns using only linear elasticity theory which correctly predicts the outcomes of several experiments.  Each element of the pattern is modeled by a ``dislocation dipole'' located at a point on a lattice, which then interacts elastically with all other dipoles in the system.  We explicitly consider a membrane with a square lattice of circular holes under uniform compression and examine the changes in morphology as it is allowed to relax in a specified direction.  
\end{abstract}

\maketitle

\section{Introduction}

It is a testament to the ingenuity of Nature that at all length scales there exists a multitude of complex self-assembled patterns.  The formation of patterns, from the textured dimples in a grain of pollen to the labyrinth of creases in a coral colony to the ridges in a fingerprint, is consistent and repeatable \cite{pollen,coral,fingerprints}; the mechanisms that drive them are far from understood.  Mastery of such processes would revolutionize the fabrication and design of novel materials with specific properties.  By harnessing elastic instabilities in elastomeric membranes \cite{nanoletters}, the once lofty goal of creating self-assembled complex patterns with long range order is now a one step closer.

When an elastomeric membrane with a square lattice of circular holes (with diameter roughly half the lattice spacing) in it is uniformly swollen or, alternatively, is compressed hydrostatically, the holes deform into a diamond plate pattern.  The order persists for upwards of $10^5$ times the original lattice spacing with only phase-slip defects, which do not change the overall symmetry of the pattern.  It is interesting to note that the same diamond plate pattern emerges from membranes with vastly different lattice spacings, ranging from $1\, \mu$m to 1 cm.  Because the same mechanism causes patterns to form over such a wide range of length scales, this technique is well suited to create devices for many fields and industries.  In particular, when a material is in the diamond plate state, it has a photonic band-gap \cite{bandgap}; hence, merely by compressing and relaxing a membrane, we can control its band structure.

This effect occurs in the highly non-linear regime of elasticity; thus, only finite element simulations using specific models of nonlinear elasticity capture the entire process of the holes collapsing \cite{finite element}.  However, some simulations merely predict the orientational order of the collapsed holes and cannot capture the details of their final shape.  With the sole assumption that each hole collapses to some elongated shape, our model uses only linear elasticity to successfully predict the orientational order in the diamond plate pattern and the herringbone pattern formed from an underlying triangular lattice.  Not only does our model shed light on the interactions in the system, it greatly facilitates the rational design of other patterns and devices.  

In the next section, we review the linear theory of elasticity and  consider a hydrostatically compressed membrane containing either a single hole or two holes. The breakdown of linear elasticity illustrates the need for non-linear analysis.  In section III, we adopt
 the spirit of the theory of cracks and show that complex behavior can be explained by recasting the problem using linear dislocation theory.  While an explicit analytic description of the shape of a collapsed hole cannot be reached through this method, each hole can be modeled as an elastically interacting distribution of parallel edge dislocations, or ``dislocation dipoles,'' whose centers are fixed to the center of their corresponding hole but are allowed to rotate freely.  In the simple case of a quartet of holes, the resulting configuration exactly reproduces the unit cell for the diamond plate pattern.  Since elastic interactions are long ranged, the collective interactions of all the holes must be included to find the true groundstate of the system-- the inclusion of which additionally stabilize the pattern.  We end with technical appendices which argue the validity of our approximation.

\section{In Which The Linear Theory of Elasticity Fails}
\subsection{Linear Elasticity in a Nutshell}
The theory of linear elasticity describes the deformation and energetics of a solid body under external force or load.  While this theory was originally developed over a century ago and has been presented time and again \cite{love,landau,green,heinbockel}, this brief tutorial will illuminate a few salient features as well as familiarize the reader to our notation.  The goal is to develop the framework for a generalization of Hooke's Law for solid three dimensional bodies.  When a solid is deformed, the displacement of every point, $\boldsymbol{x}$ is described by the vector $\boldsymbol{u}$, such that its final positions are given by $ \boldsymbol{x}'=\boldsymbol{x} + \boldsymbol{u}(\boldsymbol{x}).$  Let us first consider two points separated by $ds=\sqrt{\boldsymbol{dx} \cdot \boldsymbol{dx}}$ that are very close together.  After being deformed, their separation becomes $ds'$, where $ds '^{2}= (dx_i+du_i)^2$ using the Einstein summation convention.  By noting $du_i= (\partial u_i/ \partial x_k)dx_k$, we may rewrite this as 
\begin{eqnarray}
ds'^2 &=& ds^2 + \sum_{ik}\left( \frac{\partial u_i}{\partial x_k} + \frac{\partial u_k}{\partial x_i} +  \frac{\partial u_i}{\partial x_k} \frac{\partial u_k}{\partial x_i} \right) dx_i dx_k\nonumber\\
&=& ds^2 + 2\sum_{ik} u^L_{ik} dx_i dx_k,
\end{eqnarray}
where $u^L_{ik}$ is the Lagrangian strain tensor.  However, for small deformations we need only consider terms linear in $\partial u_i/\partial x_k$ and  the linearized strain tensor is
\begin{equation}
u_{ik} = \frac{1}{2} \left( \frac{\partial u_i}{\partial x_k} + \frac{\partial u_k}{\partial x_i} \right).
\end{equation}
In the following, we will rely upon orthogonal coordinates, $\xi_\mu$, with the diagonal metric
$ds^2=h_\a^2 d \a^2 + h_\b^2 d \b^2=\sum_\mu(h_\mu d\xi_\mu)^2$ \cite{note1}.  In general, the orthogonal coordinates are nonholonomic but the benefit of orthogonality outweighs this complication.

What is the linearized Lagrangian strain tensor in our new coordinates?  
By definition
\begin{equation}
ds'^2 -ds^2=2 u^L_{\mu \nu}h_\mu d\xi_\mu h_\nu d\xi_\nu
\end{equation}
and 
\begin{equation}
ds'^2 =\sum_i \left(dx_i +\frac{\partial u_i}{\partial x_k} dx_k\right)\left(dx_i +\frac{\partial u_i}{\partial x_j} dx_j\right)
\end{equation}
Since $\boldsymbol{u}$ is a vector, we may write it in either coordinate system:
\begin{equation}
\boldsymbol{u} = u_i \hat x_i = u_\mu \hat \xi_\mu
\end{equation}
from which we have $u_i = \sum_\mu u_\mu \hat\xi_\mu\cdot\hat x_i$.  In order
to calculate the direction cosines, we note that if $x_i=x_i(\xi_\mu)$, then
\begin{equation}
\hat \xi_\mu = \frac{1}{\sqrt{\sum_j \left(\frac{\partial x_j}{\partial \xi_\mu}\right)^2}}\sum_i\frac{\partial x_i}{\partial \xi_\mu}\hat x_i
\end{equation}
We recognize the radicand in the denominator as $h_\mu^2$ and we have
\begin{equation}
\left(dx_i +\frac{\partial u_i}{\partial x_k} dx_k\right) = \left[ \frac{\partial x_i}{\partial \xi_\mu} + 
\frac{\partial}{\partial\xi_\mu}\left(\frac{u_\rho}{h_\rho} \frac{\partial x_i}{\partial \xi_\rho}\right)\right]d\xi_\mu
\end{equation}
Using orthogonality, $\sum_i \frac{\partial x_i}{\partial\xi_\mu}\frac{\partial x_i}{\partial \xi_\nu} = h^2_\mu \delta_{\mu\nu}$,  it is straightforward to find
\begin{eqnarray}\label{strain ortho}
u_{\mu \nu} &=&  \sum_{i\rho}\left[ h_\mu \frac{\d u_\mu}{\d \xi_\nu}+h_\nu \frac{\d  u_\nu}{\d \xi_\mu}
 +\frac{\d}{\d \xi_\mu}\left(\frac{1}{h_\rho}\frac{\d x_i}{\d \xi_\rho} \right) \frac{\d x_i}{\d \xi_\nu} u_\rho  \right.\nonumber\\&&+\left. \frac{\d}{\d \xi_\nu}\left(\frac{1}{h_\rho}\frac{\d x_i}{\d \xi_\rho} \right) \frac{\d x_i}{\d \xi_\mu} u_\rho \right]/(2h_\mu h_\nu),
\end{eqnarray}
where we only sum over the repeated indices $i$ and $\rho$, not $\mu$ or $\nu$ associated with the scale factor $h$.  In the following we will adopt the Einstein summation convention and, only when there is ambiguity, will we specify which indices are to be implicitly summed.

In the spirit of Hooke's Law, there is an energy cost associated with displacing every point from its equilibrium position.  While the general form is quite complicated, to first order the energy can be constructed from the linear strain tensor.  Because the strain tensor is a symmetric rank two tensor, the only two possible scalar invariants that can be constructed are $(u_{ii})^2$ and $(u_{ik})^2$.  Because these terms are invariants of the system, they cannot depend on coordinate system.  From this we deduce the form of the energy density
\begin{equation}
\epsilon=\frac{1}{2}\left(\lambda {u_{ii}}^2 +2 \mu {u_{ik}}^2\right),
\end{equation}
where $\lambda$ and $\mu$ are Lam\'e coefficients.  Since any deformation may be written in terms of uniform or hydrostatic compression and pure shear, we rewrite the strain tensor, $u_{ik} = \frac{1}{3} \delta_{ik} u_{ll} +(u_{ik} - \frac{1}{d} \delta_{ik} u_{ll}).$  The hydrostatic compression is given by the first term, as it involves only the trace of $u_{ik}$.  And the second term describes pure shear because its trace is zero.  In terms of these quantities, the energy becomes
\begin{equation}\label{bulk}
\epsilon=  \frac{1}{2}\left(K {u_{ll}}^2 + 2 \mu \left(u_{ik} - \textstyle{\frac{1}{d} }\delta_{ik} u_{ll}\right)^2\right),
\end{equation}
where the bulk modulus $K = \lambda + \frac{1}{d} \mu$, and $\mu$ is the shear modulus.

When a body is deformed, the displaced internal elements experience forces which tend to restore them to their equilibrium positions.  The volume element bounded by the surfaces $x=x_0$,  $x=x_0+\delta x$, $y=y_0$,  $y=y_0+\delta y$, $z=z_0$, and $z=z_0+\delta z$ experiences a force along any surface with normal $\boldsymbol{n}$, $f_i=\s_{ik}n_k$, where $\s_{ik}$ is the stress tensor and
\begin{equation}
F^{\textrm{ext}}_i = \oint dA \, \s_{ik} \, n_k = \int dV \frac{\d \s_{ik}}{\d x_k}.
\end{equation}
In equilibrium, the internal stresses of the system must balance the external forces exerted upon it; thus, the equilibrium condition for the system is
\begin{equation}
 \frac{\d \s_{ik}}{\d x_k} - f^{\textrm{ext}}_i=0.
\end{equation}
In curvilinear coordinates, this entire discussion can be repeated and the stress tensor in these coordinates is merely a transformation of the stress tensor in Cartesian coordinates:
\begin{equation}\label{stress1}
\s_{\mu \nu} = \frac{1}{h_\mu}\frac{1}{h_\nu} \sum_{ik}\frac{\d x_i}{\d \xi_\mu} \frac{\d x_k}{\d \xi_\nu} \s_{ik},
\end{equation}
where, again, there is no sum over $\mu$ or $\nu$.
The equilibrium conditions become
\begin{eqnarray}\label{eq ortho}
& &\sum_{i\nu\lambda}\frac{1}{h_\nu} \frac{\d}{\d \xi_\nu} \left( \frac{1}{h_\lambda} \frac{\d x_i}{\d \xi_\lambda} \right) \left[ \frac{1}{h_\mu} \frac{\d x_i}{\d \xi_\mu} \s_{\nu \lambda}+ \frac{1}{h_\nu} \frac{\d x_i}{\d \xi_\nu} \s_{\mu \lambda} \right]\nonumber\\
& & \quad+ \frac{1}{h_\nu}\frac{\d \s_{\mu \nu}}{\d \xi_\nu} -f^{\textrm{ext}}_\mu =0.
\end{eqnarray}
where we have followed the same procedure as in the derivation of $u_{\mu\nu}$ which requires expanding the tensor in both the $\hat x_i$ and $\hat \xi_\mu$ frames.

The relation between stress and strain follows the argument in Cartesian coordinaties.
If the system is deformed an infinitesimal amount $\delta u_i$, the work done by the change in internal stresses is the force times the displacement, $W=\int dV \, (\partial \sigma_{ik}/\partial x_i) \delta u_k.$  When integrated by parts, the work is
\begin{eqnarray}
W&=& \oint dA \, \sigma_{ik} \, n_k \, \delta u_i - \int dV \sigma_{ik} \frac{\partial \delta u_i}{\partial x_k}\nonumber \\
&=& -\frac{1}{2} \int dV \sigma_{ik} \left( \frac{\partial \delta u_k}{\partial x_i} + \frac{\partial \delta u_i}{\partial x_k} \right) \nonumber\\
&=& -\int dV \sigma_{ik} \, d u_{ik}.
\end{eqnarray}
Note that the surface integral vanishes because $\sigma_{ik}=0$ at infinity.  Thus, $dE = \sigma_{ik} \, \delta u_{ik}.$  By taking the total differential of equation (\ref{bulk}), $dE = K \, u_{ll} \, du_{ll} + 2 \mu (u_{ik}- \frac{1}{d} \delta_{ik}u_{ll})d (u_{ik} - \frac{1}{d} \delta_{ik} u_{ll} ) =\left (K \, u_{ll} \, \delta_{ik} + 2 \mu (u_{ik} - \frac{1}{d} \delta_{ik} u_{ll}) \right) du_{ik}$, we can rewrite the stress tensor in terms of the strain tensor, $
\sigma_{ik} = K \delta_{ik} u_{ll} -2 \mu \left( u_{ik} -\textstyle{\frac{1}{d}} \delta_{ik} u_{ll} \right),$ or conversely, the strain in terms of the stress, $u_{ik} = \textstyle{\frac{1}{d^2 K}}\delta_{ik} \sigma_{ll}+ \textstyle{\frac{1}{2 \mu}}\left(\sigma_{ik} - \textstyle{\frac{1}{d}} \delta_{ik} \sigma_{ll} \right).$

Similarly, it is instructive to consider a homogeneous deformation, which has constant stain tensor everywhere in the volume.  For a $d$-dimensional solid, uniform pressure is applied to the faces with normals in the $\pm \hat{z}$ directions.  In 3-dimensions, for example, we consider the simple compression of a rod.  This implies that $\sigma_{zi} n_i=p,$ or $\sigma_{zz}=p$.  Thus, all off diagonal components of the strain tensor are zero, and the diagonal components are $u_{ll} =p(\frac{1}{dK} -\frac{1}{2\mu} )/d,$ for all $l \ne z,$ and $u_{zz} = p ( \frac{1}{dK} + \frac{d-1}{\mu} )/d$.  The relative longitudinal compression is given by $u_{zz} = p/Y_d,$ where 
\begin{equation}
Y_d=\frac{2d^2K\mu}{2\mu+(d^2-d)K}
\end{equation}
 is the $d$-dimensional Young's modulus.  The ratio of transverse extension to longitudinal compression is given by the Poisson ratio $\g = -u_{ll}/u_{zz}=(dK-2\mu)/((d^2-d)K+2\mu)$.  Conversely, the bulk and shear moduli written in terms of the Young's modulus and Poisson ratio are, respectively, $K= \frac{1}{d}Y_d/(1-(d-1)\g)$ and $\mu = \frac{1}{2}Y_d/(1+\g)$.  The stress tensor is given in terms of the strain tensor by,
\begin{equation}\label{stress from strain}
\sigma_{ik} = \frac{Y_d}{1+\g} \left(u_{ik} + \frac{ \g}{1-(d-1)\g} \delta_{ik} u_{ll}\right),
\end{equation}
and the converse by,
\begin{equation}\label{strain from stress}
u_{ik} = \frac{1}{Y_d} \left( ( 1+\g) \sigma_{ik} - \g \delta_{ik} \sigma_{ll} \right).
\end{equation}
The conventional form of the energy is given in terms of $Y_d$ and $\g,$
\begin{equation}
E= \frac{Y_d}{2(1+\g)}\int d^dx \left({u_{ik}}^2+\frac{\g}{1-(d-1)\g}{u_{ll}}^2 \right).
\end{equation}
Now that we have derived the relations between the stress and strain tensors, the equilibrium condition may be recast in terms of the displacement vector,
\begin{equation}
\frac{Y_d}{2 (1+ \g)} \left(\frac{\partial^2 u_i}{\partial x_k^2}+\frac{1-(d-3) \g}{1-(d-1) \g} \frac{ \partial^2 u_l}{\partial x_i \partial x_l}\right)-f_i =0.
\end{equation}
If the force only acts through the surface, then $f_i$ vanishes in the bulk and we recover the result that 
$\nabla^2\nabla\cdot\boldsymbol{u}=0$ from which it follows that $\nabla^4 \boldsymbol{u}=0$.

\subsection{Elasticity in \textit{Flatland}\cite{flatland}}
Our ultimate goal is to study the effects of uniform tension on a thin sheet of elastic material with a square lattice of circular holes cut in it, as it models an elastic sheet that is uniformly swollen.  In general, the resulting deformations are constant across the thickness of the sheet and may be considered to be purely longitudinal.  This is because forces act primarily in the plane of the film, yielding the boundary condition $\sigma_{ik}n_k=0$.  Henceforth, we will only consider two dimensional systems with planar deformations.

Let us pause for a minute to derive the equilibrium conditions of this system.  Since the deformations are constant throughout the thickness of the sheet, we may assume that $u_{zz}=0,$ and thus $\sigma_{zz} = \sigma_{iz} =0$.  We are left with the two equilibrium equations $h \partial \sigma_{ik}/\partial x_k =-p_i,$ where $i,k=x,y$, and $h$ is the thickness of the film.  When no external body forces are present the equations of equilibrium reduce to $\partial \sigma_{ik}/\partial x_k=0$ or,
\begin{equation}\label{airy cartesian}
\frac{\partial \sigma_{xx}}{\partial x} + \frac{\partial \s_{xy}}{\partial y} = 0, \ \ \frac{\partial \sigma_{xy}}{\partial x} + \frac{\partial \s_{yy}}{\partial y}= 0.
\end{equation}
Viewing these both as equations of the form $\nabla\cdot\boldsymbol{A}=0$, we know that
$\s_{xi} = \epsilon_{ik}\partial_k \phi_y$ and $\s_{jy}=\epsilon_{jk}\partial_k\phi_x$ and it follows that $\partial_x\phi_y+\partial_y\phi_x=\s_{xy}-\s_{xy}=0$ (where $\epsilon_{i k}$ is the totally antisymmetric tensor).    Thus $\phi_m=\epsilon_{mn}\partial_n\chi$
for some scalar $\chi$, known as the Airy stress function.  We thus have $\s_{ik}=\epsilon_{im}\epsilon_{kn}\partial_m\partial_n\chi$.  Moreover, since $u_{ik}=(\partial_iu_k +\partial_ku_i)/2$, we have 
$\epsilon_{im}\epsilon_{kn}\partial_i\partial_ku_{mn}=0$.  The relation between stress and strain, (\ref{stress from strain}) implies that
\begin{equation} 
\epsilon_{im}\epsilon_{kn}\frac{\partial^2\s_{mn} }{\partial x_i\partial x_k}= \frac{Y_d}{1+\g}\left(0 + \frac{\g}{1-(d-1)\g}\nabla^2\nabla\cdot\boldsymbol{u}\right)=0
\end{equation}
As a result, 
we deduce that the stress function satisfies the biharmonic equation, $\nabla^4 \chi=0.$  In orthogonal coordinates, we recast the components of the stress tensor in terms of the Airy stress function by taking advantage of Eq. (\ref{stress1}) and reexpressing $\sigma_{ij}$ in terms of $\sigma_{\mu\nu}$,
\begin{equation}\label{airy ortho}
\s_{\mu \nu}= \sum_{\rho\lambda}\epsilon_{\mu \lambda} \frac{1}{h_\lambda} \left[ \epsilon_{\nu \rho} \frac{\d}{\d \xi_\lambda} \Big(\frac{1}{h_\rho} \frac{\d \chi}{\d \xi_\rho} \Big) + \epsilon_{\mu \rho} \frac{1}{h_\mu h_\nu} \frac{\d h_\rho}{\d \xi_\mu} \frac{\d \chi}{\d \xi_\nu} \right],
\end{equation}
where $\epsilon_{\a \b}=1$ because $\{ \hat{\boldsymbol{\a}}, \hat{\boldsymbol{\b}}\}$ is a right-handed orthonormal basis.  Equations (\ref{airy ortho}) are solutions to the equilibrium equations, Eq. (\ref{eq ortho}), and the Airy stress function solves the biharmonic equation in {\sl orthogonal} coordinates.

\subsubsection{Fixing a Hole (Demo)}
The simplest system to study is an infinite elastic sheet with a circular hole, of radius $R,$ cut in it under uniform tension $P\bf{\hat{x}}$.  This problem naturally lends itself to polar coordinates, for which the equations for the stress function become,
\begin{eqnarray}
 \sigma_{rr}& =& \frac{1}{r} \frac{\partial \chi}{\partial r} + \frac{1}{r^2} \frac{\partial ^2 \chi}{\partial \phi^2}, \ \ \sigma_{\phi\phi} = \frac{\partial ^2 \chi}{\partial \phi^2},  \nonumber\\
\sigma_{r\phi} &=& -\frac{\partial}{\partial r} \left(\frac{1}{r}\frac{\partial \chi}{\partial \phi}\right). 
\end{eqnarray}
A standard procedure for solving such problems is to solve first for the deformation of a continuous sheet under the proper forces. Secondly, we solve a for a second stress function respecting the symmetry broken by the force with boundary conditions $\sigma_{ik}(r=\infty)=0$.  The final stress function is given by the sum of the two stress functions, where the matching condition is given by the stress free boundary condition at the edge of the hole.\\
\indent The components of the stress tensor for a continuous elastic sheet under uniform tension $P \bf{\hat{x}}$ are $\sigma_{xx}^\0=P$ and $\sigma_{yy}^\0=\sigma_{xy}^\0=0$, which, by integrating Eqs. (\ref{airy cartesian}) yield the stress function
\begin{equation}
\chi^\0=P y^2/2=Pr^2 (\sin^2 \phi) /2=Pr^2(1-\cos 2\phi)/4
\end{equation}
from which it follows that the components of the stress tensor are $\sigma_{rr}^\0(r) = P(1+ \cos 2 \phi)/2$, $\sigma_{\phi\phi}^\0(r)=P(1- \cos 2 \phi)/2$, and $\sigma_{r \phi}^\0(r)=P( \sin 2 \phi) /2$.  Clearly, rotational symmetry is broken in the $\bf{\hat{x}}$-direction.

In order for the second stress function to respect the broken symmetry of the system, it must have the form $\chi^\1=f(r)+g(r) \cos 2 \phi.$  Since the stress function satisfies the biharmonic equation, we can easily integrate to find 
\begin{eqnarray}
f(r) &=& a r^2 \log r+b r^2 + c \log r\nonumber\\
g(r) &=& s r^2 + t r^4 + u/r^2+v
\end{eqnarray}
The first boundary conditions $\sigma^\1_{ik} (r=\infty)=0$ dictate that $a=b=s=t=0$, leaving 
\begin{eqnarray}
\sigma_{rr}^\1(r) &=& c/r^2 - \left(6 u/r^4+4 v/r^2\right)\cos 2 \phi\nonumber\\
\sigma_{\phi\phi}^\1(r)&=& -c/r^2 +6(u/r^4) \cos 2 \phi\nonumber\\
\sigma_{r\phi}^\1(r)& =& -\left(6 u/r^4+4 v/r^2 \right) \sin 2 \phi
\end{eqnarray}
Using the final boundary conditions, $\sigma^\1_{\mu\nu}(R)=-\sigma^\0_{r\phi}(R)$ the remaining constants are found to be $c=-P R^2/2$, $u=-P R^4/4$, and $v=P R^2/2$.  Note that even though there are only two equations for three unknowns, this system is not underdetermined because the constant $c$ cannot depend on $\phi$.  Assembling this, the components of the stress tensor are given by
\begin{eqnarray}
\sigma_{rr}(r) &=& \frac{P}{2}\left[1-\frac{R^2}{r^2}+\left(1-\frac{4 R^2}{r^2}+\frac{3 R^4}{r^4} \right)\cos 2 \phi\right],  \nonumber\\
\sigma_{\phi\phi}(r)&=& \frac{P}{2}\left[1+\frac{R^2}{r^2}-\left(1+ \frac{3 R^4}{r^4} \right)\cos 2 \phi \right]\nonumber\\
\sigma_{r\phi}(r) &=& -\frac{P}{2} \left(1+ \frac{2 r^2}{r^2} -\frac{3 R^4}{r^4}\right) \sin 2 \phi .
\end{eqnarray}
which may be rewritten, using equation (\ref{strain from stress}), as components of the strain tensor
\begin{eqnarray}
u_{rr}&=&\frac{1}{Y_2}(\sigma_{rr} -\g \sigma_{\phi\phi})\nonumber\\
u_{\phi\phi}&=& \frac{1}{Y_2}(\sigma_{\phi\phi} -\g \sigma_{rr})\nonumber\\
u_{r\phi} &=& \frac{1+\g}{Y_2}\sigma_{r \phi}.
\end{eqnarray}

Recall, the strain tensor is the relative displacement of every element from its equilibrium position.  In polar coordinates we have $u_{rr}=\partial u_r/\partial r$, $u_{\phi\phi} = (\partial u_\phi/\partial \phi)/r+u_r/r,$ and $2 u_{r\phi} = \partial u_\phi/\partial r+(\partial u_r/\partial \phi)-u_\phi/r$, from which it follows that
the displacement vector is 
\begin{widetext}
\begin{eqnarray}
u_r&=& \frac{P}{2Y_2} \left[(1-\g)r+ \frac{(1+\g) R^2}{r} + \left((1+\g)\left(r-\frac{R^4}{r^3} \right) +\frac{4 R^2}{r}\right) \cos 2\phi \right]\nonumber\\
u_\phi&=& -\frac{P}{2Y_2}\frac{(R^2+r^2)^2+\g (R^2-r^2)^2}{r^3} \sin 2 \phi.
\end{eqnarray}
\end{widetext}

However, if we repeat the above process for an infinite sheet under hydrostatic compression (or expansion), to linear order the rotational symmetry of this system is not broken.  Consider an annulus of inner radius $R_0$ and outer radius $R_1$ under uniform hydrostatic compression with boundary conditions given by $\sigma_{rr}^\0=-P$ and $\sigma_{\phi\phi}^\0=\sigma_{r\phi}^\0=0$.  Clearly, the displacements are purely radial, and we need only solve $\nabla^4 \boldsymbol{u}=0$, subject to the boundary conditions $\s_{rr}(r=R_1)=-P$ and $\s_{rr}(r=R_0)=0$.  The displacements are given by $u_r(r)=a/r+br$; from which it follows $\s_{rr}=\frac{Y_2}{1-\g^2}\left[-(1-\g)a/r^2+(1+\g)b\right].$  The boundary conditions determine the values of the constants $a=-\frac{P}{Y_2}(1+\g)\frac{R_0^2R_1^2}{R_1^2-R_0^2}$ and $b=-\frac{P}{Y_2}(1-\g)\frac{R_1^2}{R_1^2-R_0^2}$.  The solution to an annulus under hydrostatic compression is
\begin{equation}
u_r(r)=\frac{P}{Y_2}\frac{R_1^2}{R_1^2-R_0^2}\left((1+\g)\frac{R_0^2}{r}+(1-\g)r\right),
\end{equation}
with the components of the stress tensor given by
\begin{eqnarray}
\s_{rr}(r)&=&-P \frac{R_1^2}{R_1^2+R_0^2}\frac{r^2-R_0^2}{r^2}\nonumber\\
\s_{\phi\phi}(r)&=&-P \frac{R_1^2}{R_1^2+R_0^2}\frac{r^2+R_0^2}{r^2}\nonumber\\
\s_{r\phi}(r)&=&0.
\end{eqnarray}

In the case of a finite sheet or a pipe under hydrostatic compression, there is the well known von Mises buckling instability at a critical pressure where the circular hole deforms into an ellipse whose major axis is chosen at random \cite{von mises}.  This critical pressure scales with the ratio of the system size to the hole radius and, thus, diverges for large systems.  The system we are studying, on the other hand, has an underlying lattice which breaks the rotational symmetry of each hole.  A superposition of the above solutions would not account for the interaction between holes.

\subsubsection{Fixing a Hole (Take 2)}
Understanding the elastic interaction between holes in an elastic sheet was, during the first half of the last century, the subject of much research \cite{jeffery,howland1,howland2,green,stevenson,ling}.  Most of which was dedicated to finding the maximum stress felt along the perimeter of each hole.  While linear elasticity may provide reasonable solutions to such analysis, we will demonstrate that linear theory breaks down upon further investigation.

\begin{figure}[!h]
\begin{center}
\includegraphics[width=3.5in]{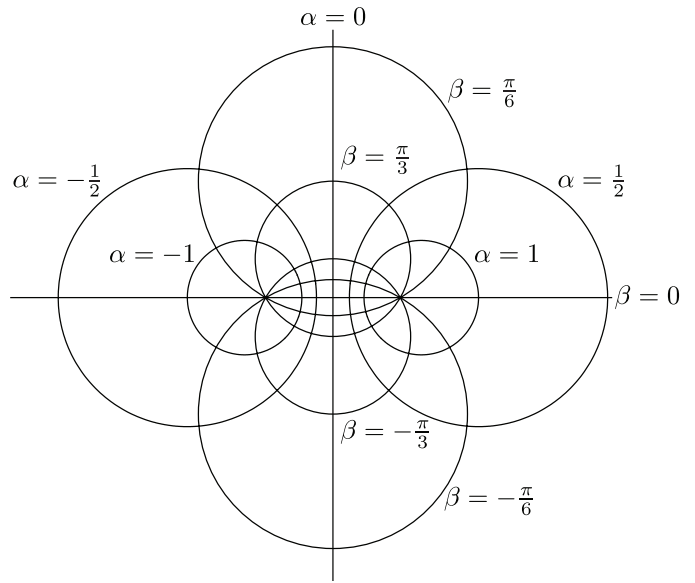}
\caption{Curves of constant $\a$ and $\b$ are circles in bipolar coordinates.}
\label{bipolar coordinates}
\end{center}
\end{figure}

\begin{figure}[!h]\label{bipolar pics}
\begin{center}{
\includegraphics[width=3.75in]{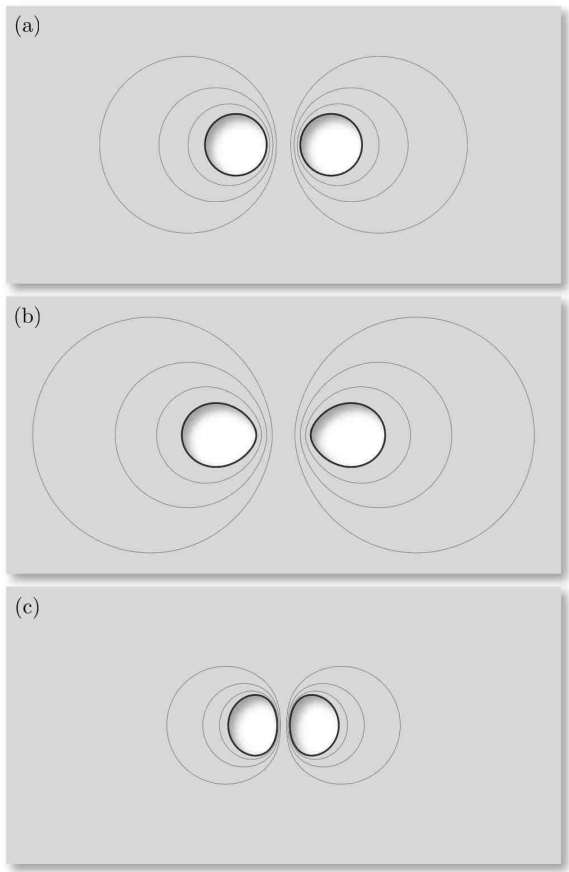}
\caption{An elastic sheet with two circular holes cut out (a) is subjected to uniform tension (b) and compression (c).  The dark blue curves are the the boundary of the holes.  The other curves show deformations of the circles in (a) to aid the eye.}
\label{two holes}}
\end{center}
\end{figure}

The simplest system accounting for the interaction between holes is an infinite elastic sheet containing two holes of radius $R$, whose centers are separated by distance $2 d$.  This sheet is then subjected to uniform tension $P$.  The analysis of this system will closely follow that of Ling \cite{ling}.  Bipolar coordinates, defined by
\begin{equation}
x= \frac{a \sinh \a}{\cosh \a-\cos \b}, \quad y=\frac{a \sin \b}{\cosh \a-\cos \b},
\end{equation}
for $\b \in [0,2\pi), \a \in (-\infty,\infty),$ are the natural choice for this problem, as lines of constant $\a$ or $\b$ are circles in the $xy$-plane defined by $x^2 +(y-a \cot \b)^2 = a^2 \csc^2 \b$ and $(x-a \coth \a)^2+y^2=a^2 \csch^2 \a$.  The system of two equal holes corresponds to $\a=\pm s$,  $s=\cosh^{-1}(d/R)$, and $a^2=d^2-R^2$.  When the system is under uniform tension $P$, the components of the stress tensor are $\s_{xx}=P$, $\s_{yy}=P$, and $\s_{xy}=0$.  By integrating the Eqs. (\ref{airy cartesian}), the stress function for an infinite system under uniform tension is
\begin{equation}\label{airy bipolar 0}
\chi^\0=\frac{P}{2}(x^2+y^2)= \frac{P a^2}{2} \frac{\cosh \a + \cos \b}{\cosh \a - \cos \b}.
\end{equation}
Using the method outlined in the previous section, we undertake the tedious calculation, detailed in Appendix \ref{orthogonal}, to find the equilibrium configuration of this system.  The results for uniform compression and tension are displayed in FIG. \ref{bipolar pics}.  Upon further analysis of the compressed system, large enough values of $P$ yield overlapping solutions for the displacement vectors, signaling the breakdown of the linear theory.

\section{The Linear Theory of Nonlinear Elasticity}
Even were there no instability in the linear theory of two elastic holes, the sheer complexity of the equations would make calculations of increasing numbers of holes a nearly impossible task, and understanding the mechanism by which the holes collapse and the shapes they form requires a nonlinear theory of elasticity.  Thus, we turn to the theory of cracks for inspiration.  In the linear theory of elasticity, cracks can be described by a continuous distribution of parallel dislocations \cite{landau,hirth}.  The stresses in a body due to a crack are the same as the stresses in an isotropic body with a distribution of dislocations with the same height profile as that of the crack.  As a first approximation, we model each of the collapsed holes as a pair of oppositely charged dislocations, known as a dislocation dipole \cite{chaikin}.  This formalism allows us to recover the same physics by the simple numerical minimization of algebraic equations, once described by a complex system of coupled differential equations.

\subsection{Filling a Crack}

\begin{figure}[!tbh]
\begin{center}
\includegraphics[width=3.25in]{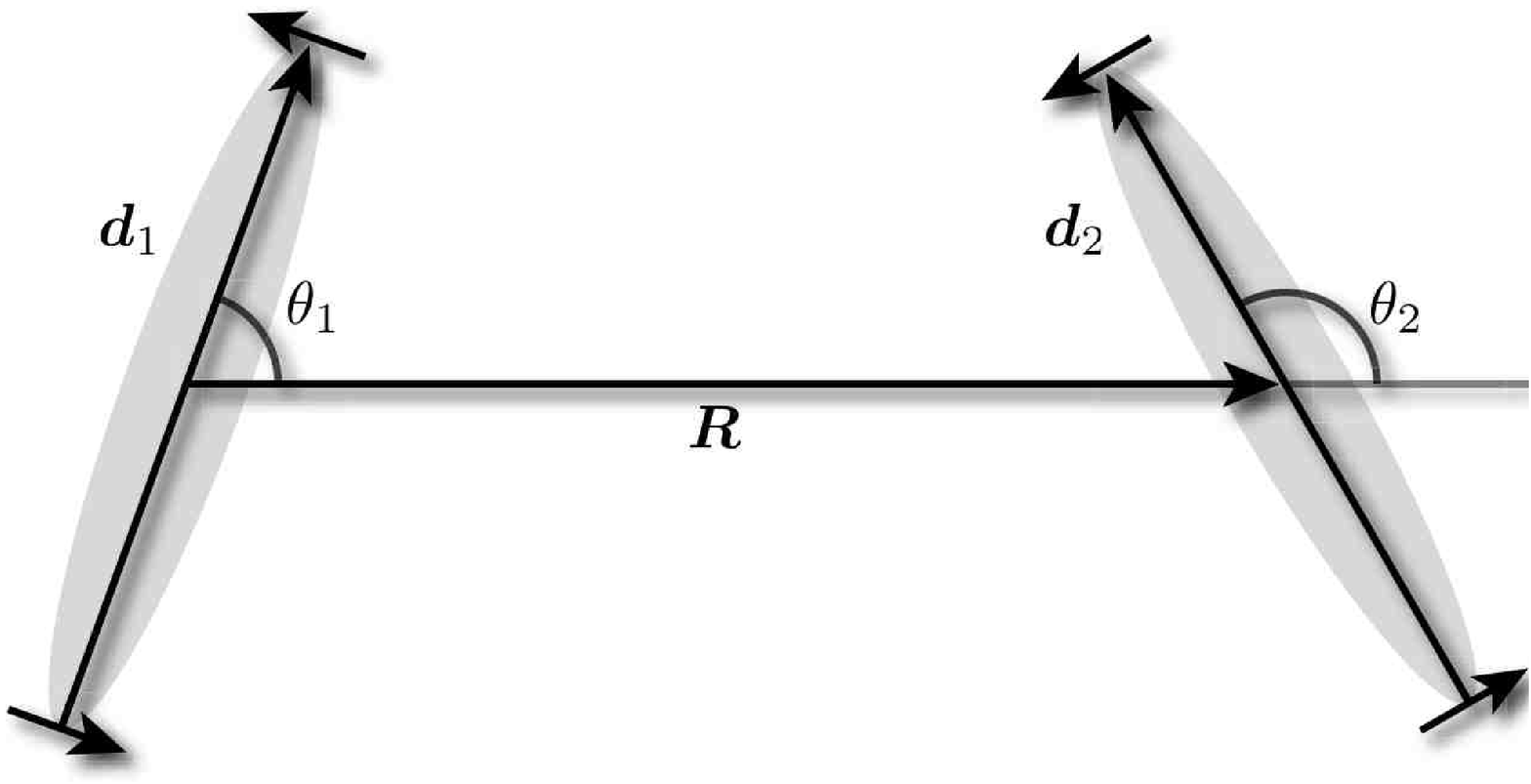}
\caption{Two dipoles of strengths $\boldsymbol{d}_1$ and $\boldsymbol{d}_2$ are separated by $\boldsymbol{R}$.}
\label{dipole interaction}
\end{center}
\end{figure}

Our model system consists of an isotropic solid where each collapsed hole is represented by a thin line of material that has been taken out of the system, or a dislocation dipole.  The Burgers vector for a dislocation dipole of strength $b$ with dipole vector $\boldsymbol{d}$ located at $\boldsymbol{r}$ is 
\begin{equation}
\boldsymbol{b}\left(\boldsymbol{x}\right) = \hat{\boldsymbol{z}} \times \hat{\boldsymbol{d}} \,b \left[ -\delta^2\left( \boldsymbol{x}-\frac{\boldsymbol{d}}{2}- \boldsymbol{r}\right)+\delta^2\left( \boldsymbol{x}+\frac{\boldsymbol{d}}{2}- \boldsymbol{r}\right)\right],
\end{equation}
or in Fourier space,
\begin{eqnarray}
\boldsymbol{b}\left(\boldsymbol{q}\right) &=& 2ib\hat{\boldsymbol{z}} \times \hat{\boldsymbol{d}}\,e^{i q r \cos \theta}\sin\left[\frac{qd}{2}\cos(\theta-\theta_0)\right]\nonumber\\
&\approx&ib q\hat{\boldsymbol{z}} \times \boldsymbol{d}\, e^{i q r \cos \theta}\cos(\theta-\theta_0).
\end{eqnarray}
where $\theta$ is angle of $\boldsymbol{q}$ and $\theta_0$ is the direction of the dipole and we have, in the spirit of the dipole approximation, taken the lowest order term in $d=\vert\boldsymbol{d}\vert$.  

The interaction energy of two dipoles $\boldsymbol{d}_1$ and $\boldsymbol{d}_2$ both of strength $b$ separated by $\boldsymbol{R}$ is given by
\begin{widetext}
\begin{eqnarray}\label{burgers integral}
E=\frac{Y_2 b^2 d_1 d_2}{(2\pi)^2} \int d^2 q\frac{\left[\boldsymbol{q} \times ( \hat{\boldsymbol{z}} \times \hat{\boldsymbol{d}}_1 )\right] \cdot \left[\boldsymbol{q}\times ( \hat{\boldsymbol{z}} \times \hat{\boldsymbol{d}}_2 )\right]}{q^4} \left(i q \cos(\th-\th_1)\right)\left(-i q e^{-iq R \cos \th} \cos(\th-\th_2)\right),
\end{eqnarray}
where $\boldsymbol{d}_1$ is at the origin.  After carrying out the integration (see Appendix \ref{dipole}), the pairwise interaction between two dislocation dipoles is 
\begin{equation}\label{dipole energy}
E_{int}=-\frac{Y_2}{\pi} \frac{b^2 d_1 d_2}{R^2} \left( \cos(\th_1+\th_2)\sin \th_1 \sin \th_2+ \frac{1}{4} \right).
\end{equation}
\end{widetext}
Note that the interaction energy is invariant under $\th_1 \rightarrow \th_1+\pi$ and $\th_2\rightarrow \th_2 + \pi$, which reaffirms each collapsed hole is represented by a line, not a vector.  The total interaction energy of an array of dislocation dipoles is merely a sum of all pairwise interactions, because we are using linear theory.  The centers of the initial holes set the position of each dislocation dipole, but they are allowed to rotate freely.  The equilibrium state minimizes the free energy over the angle each dipole makes with respect to a fixed axis.

\subsubsection{The $2\times 2$ Diamond Plate Plaquette}
The simplest case consists of four holes located at $\{\pm a/2, \pm a/2\}$.  Because all four holes have the same radius, their dipole vectors should have the same magnitude.  We minimize the energy functional, composed of the sum of six pairwise terms,
\begin{widetext}
\begin{eqnarray}
E_{2\times2}&=&-\frac{Y_2 b^2 d^2}{\pi a^2}\left[\cos(\th_1+\th_2)\sin\th_1\sin\th_2+\cos(\th_3+\th_4)\sin\th_3\sin\th_4+\cos(\th_1+\th_3-\pi)\sin(\th_1-\textstyle{\frac{\pi}{2}})\sin(\th_3-\textstyle{\frac{\pi}{2}})\right.\nonumber\\
&+& \cos(\th_2+\th_4-\pi)\sin(\th_2-\textstyle{\frac{\pi}{2}})\sin(\th_4-\textstyle{\frac{\pi}{2}})+\frac{1}{2}\cos(\th_1+\th_4-\textstyle{\frac{\pi}{2}})\sin(\th_1-\textstyle{\frac{\pi}{4}})\sin(\th_4-\textstyle{\frac{\pi}{4}})\nonumber\\
&+& \left.\textstyle \frac{1}{2}\cos(\th_2+\th_3-\textstyle{\frac{3\pi}{2}})\sin(\th_2-\textstyle{\frac{3\pi}{4}})\sin(\th_3-\textstyle{\frac{3\pi}{4}})\right],
\end{eqnarray}
\end{widetext}
over each of the angles, which are measured with respect to the $x$-axis.  Minimizing with respect to the four angles we find $\th_1=\th_4$ and $\th_2=\th_3$ by symmetry and

\begin{eqnarray}\label{min 2x2}
\sin 4\th_1-\cos 2\th_1-4 \sin2(\th_1+\th_2)&=&0\nonumber\\
\sin 4\th_2+\cos 2\th_2-4 \sin2(\th_1+\th_2)&=&0.
\end{eqnarray}
The minimum is
\begin{equation}
\th_1=\frac{1}{2}\sin^{-1}\left(\frac{1}{10}\right)=\th_2-\frac{\pi}{2}.
\end{equation}
While one might have postulated that the lowest energy configuration would have $\th_1=0$ and $\th_2=\pi/2$, it turns out that the energy is slightly lowered if these angles are slightly shifted. This is due to the finite size of our system -- as we shall see, for larger systems the dipoles align along the crystal axes and there is a boundary effect which distorts the dipole directions at the edges.  

\subsubsection{We Had to Count Them All: $n \times n$ Systems of Holes}
\begin{figure}[!ht]
\begin{center}
\includegraphics[width=3.25in]{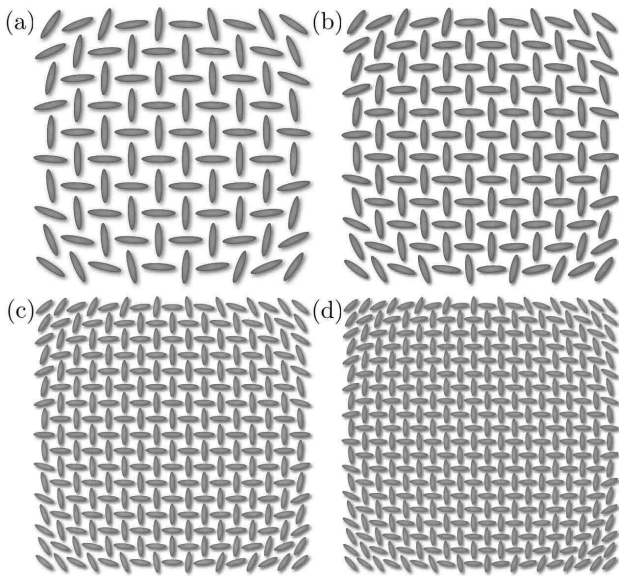}
\caption{The groundstate orientations for square lattices of $10\times10$ (a), $13\times13$ (b), $17\times17$ (c), and $20\times20$ (d).}
\label{nxn}
\end{center}
\end{figure}

One might wonder how we can study larger and larger systems since the interaction only falls off
as $1/R^2$.  Because the interactions is between dipoles, the interaction energy at large distances
decreases because the angle of the dipoles rotates around the circle.  Indeed, consider the interaction
of a single dipole at the origin with $N^2-1$ other dipoles in an $Na\times Na$ lattice.  The angular dependence in (\ref{dipole energy}) will wash out the power law if the dipoles rotate through $2\pi$ uniformly.  As we will see in the next section, this is precisely what happens as shown in Fig. \ref{nxn}.

For these larger arrays of holes, the sheer number of coupled equations makes it impractical to find solutions by hand, and we turn to numerical methods to find the orientations of the ground state of each lattice.
We find that the diamond plate order of the $2\times2$ plaquette persists for larger and larger samples with increasing alignment along  the lattice directions.

\begin{figure}[!h]
\begin{center}
\includegraphics[width=1.5in]{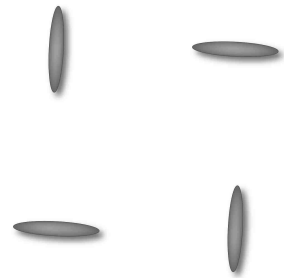}
\caption{The groundstate orientation of the $2\times2$ plaquette.}
\label{2x2}
\end{center}
\end{figure}

\subsection{Stretching the Lattice}

\begin{figure}[!ht]
\begin{center}
\includegraphics[width=3.5in]{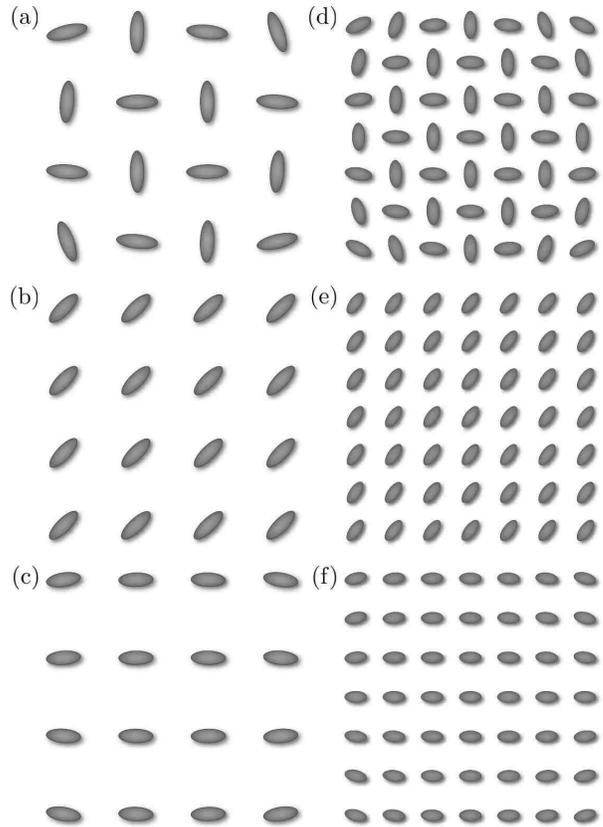}
\caption{The $4\times4$ and $7\times7$ system of holes are stretched by tension $T\boldsymbol{\hat{x}}$.  For small tensions (in (a) and (d) $T=0.1 \frac{Y_2}{\pi}\frac{b^2 d^2}{a^2}$), the diamond plate groundstate is only slightly perturbed.  Whereas for large tensions (in (b), (c), (e), and (f) $T=2 \frac{Y_2}{\pi}\frac{b^2 d^2}{a^2}$), the holes align along the $\boldsymbol{\hat{x}}$-axis, the direction of stretching.  In (b) and (e) the angle of stretching is $\th_{\rm{Stretch}}=\pi/4$ from horizontal.}
\label{stretching}
\end{center}
\end{figure}

In experimental systems \cite{nanoletters}, the elastic sheet was stretched in a specific direction before allowing it to swell, leading to a background stress $\s_{xx}=T \cos \f,$ $\s_{yy}=T \sin \f$.  To calculate the coupling energy between the stretching and the dipole angle, we will rotate the system, such that the 
$\boldsymbol{x}$-axis is defined by the direction of stretching, or $\s_{xx}=T$ and the dipole is located at the origin, or 
\begin{equation}
\boldsymbol{b}(\boldsymbol{q})=ibq \boldsymbol{\hat{z}}\times \boldsymbol{d} \cos(\th-\th_0+\f).
\end{equation}
It is most appropriate to use the energy functional $E=\frac{1}{2 Y_2} \int d^2x \left(\nabla^2 \chi\right)^2$.  In Fourier space, the dipole term is given by $\nabla^2 \chi=i \epsilon_{ik} \frac{q_k}{q^2}b_i(q)$ \cite{chaikin} and the stretching term by $\nabla^2 \chi=T\frac{\delta(q)}{q}\delta(\th)$.  The energy is
\begin{eqnarray}
E&=&\frac{1}{2 Y_2} \int \frac{d^2 q}{(2 \pi)^2} \left(T\frac{\delta(q)}{q}\delta(\th)-b d \cos^2(\th-\th_0+\f)\right)\nonumber\\
&\times& \left(-T\frac{\delta(q)}{q}\delta(\th)+b d \cos^2(\th-\th_0+\f)\right),
\end{eqnarray}
from which the coupling energy is
\begin{eqnarray}
E_{\rm{stretch}}&=&-\frac{Tb d}{Y_2(2\pi)^2}\int q dq \int d\th \frac{\delta(q)}{q}\delta(\th) \cos^2(\th-\th_0+\f)\nonumber\\
&=&-\frac{Tbd}{Y_2 (2\pi)^2}\cos^2(\f-\th_0)
\end{eqnarray}
The new term causing the dislocation dipoles to align with the direction of stretching competes with original interaction energy, favoring the diamond plate pattern.  Following the same minimization procedure as before, we find that for small tensions the diamond plate pattern is only slightly perturbed, and for large tensions, the dislocation dipoles align along the direction of stretching, which may be seen in Fig.\ref{stretching}.

\section{Conclusion}
We have created a model system for an elastic sheet with a square lattice circular holes cut out of it.  When the sheet is swollen, or, equivalently, subjected to uniform tension, the holes snap shut.  Their major axes align into a diamond plate pattern with long ranged order.  While other methods of calculating the orientational order of the holes rely upon nonlinear elasticity theory and finite element simulations, we use simple linear elasticity theory to obtain the same results.  Our system also corroborates experimental results of the sheet under external forces.  This method may easily extended to holes on other lattices.  It is difficult to extend it to an infinite lattice, since the minimization would then be over an infinite number of angles.  An Ewald type summation may be used for an infinite system whose unit cell is the $2\times2$ plaquette.  However, this is unlikely to lead to new insight, as the diamond plate order is clearly maintained for large systems.  Moreover, as we show in the appendix, the dipole interactions are the dominant terms even for more general elliptical holes.  
\acknowledgments
We thank G.P. Alexander, B.G. Chen, C.D. Modes, S. Yang, and Y. Zhang for useful discussions.  This work was supported by NSF MRSEC Grant DMR05-20020.

\appendix

\section{Linear Elasticity in Bipolar Coordinates}\label{orthogonal}
\subsection{The Basics}
For clarity's sake, we take a moment to explicitly write out the useful equations in bipolar coordinates.  From equations (\ref{strain ortho}), we simply read off the components of the strain tensor
\begin{eqnarray}\label{strain bipolar}
u_{\a\a} &=& \frac{1}{h} \frac{\d u_\a}{\d \a} + \frac{1}{h^2} \frac{\d h}{\d \b} u_\b, \quad u_{\b\b} = \frac{1}{h} \frac{\d u_\b}{\d \b}+ \frac{1}{h^2}\frac{\d h}{\d \a} u_\a, \nonumber\\
2 u_{\a\b}&=& \frac{\d}{\d \a} \left(\frac{u_\b}{h}\right) + \frac{\d}{\d\b} \left(\frac{u_\a}{h}\right).
\end{eqnarray}
The stress tensor may be written as a function of the Airy stress function $\chi$, from equation (\ref{airy ortho})
\begin{eqnarray}\label{stress bipolar}
\s_{\a\a}&=& \frac{1}{h} \frac{\d }{\d \b} \left(\frac{1}{h} \frac{\d \chi}{\d \b} \right) + \frac{1}{h^3} \frac{\d h}{\d \a} \frac{\d \chi}{\d \a},\nonumber\\
\s_{\b\b}&=& \frac{1}{h} \frac{\d }{\d \a} \left(\frac{1}{h} \frac{\d \chi}{\d \a} \right) + \frac{1}{h^3} \frac{\d h}{\d \b} \frac{\d \chi}{\d \b},\nonumber\\
\s_{\a\b}&=&- \frac{1}{h^2} \left( \frac{\d^2 \chi}{\d \a \d \b} - \frac{1}{h} \frac{\d h}{\d \b} \frac{\d \chi}{\d \a}- \frac{1}{h} \frac{\d h}{\d \a} \frac{\d \chi}{\d \b} \right)\nonumber\\
&=&-\frac{1}{2}\left[ \frac{\d}{\d \a}\left(\frac{1}{h^2} \frac{\d\chi}{\d \b} \right)+\frac{\d}{\d \b}\left(\frac{1}{h^2} \frac{\d\chi}{\d \a} \right)\right].
\end{eqnarray}
or in terms of the relative displacement vector, from Eq. (\ref{stress from strain})
\begin{eqnarray}
\s_{\a\a}&=& \frac{Y_2}{1-\s^2}\left( u_{\a\a}+\s u_{\b\b}\right),\nonumber\\
 \s_{\b\b}&=& \frac{Y_2}{1-\s^2}\left( u_{\b\b}+\s u_{\a\a}\right),\nonumber\\
2\s_{\a\b}&=& \frac{Y_2}{1+\s}u_{\a\b}.
\end{eqnarray}
At first glance, it seems an insurmountable goal to solve the differential equations for the displacements in terms of the Airy stress function.  However, following \cite{jeffery}, they become much more tractable if one considers the terms
\begin{eqnarray}\label{saa+sbb}
\s_{\a\a}+\s_{\b\b}&=&\frac{Y_2}{1-\s}\frac{1}{h^2}\left( \frac{\d(h u_\a)}{\d \a}+\frac{\d(h u_\b)}{\d\b}\right)\nonumber\\
&=&\frac{1}{h^2}\left(\frac{\d^2\chi}{\d\a^2}+\frac{\d^2 \chi}{\d\b^2}\right),
\end{eqnarray}
\begin{eqnarray}\label{saa-sbb}
\s_{\a\a}-\s_{\b\b}&=&\frac{Y_2}{1+\s} \left( \frac{\d}{\d\a} \left(\frac{u_\a}{h}\right) -\frac{\d}{\d\b} \left(\frac{u_\b}{h}\right) \right) \nonumber\\
&=&-\frac{\d}{\d\a} \left( \frac{1}{h^2}\frac{\d\chi}{\d\a}\right)+\frac{\d}{\d\b} \left( \frac{1}{h^2}\frac{\d\chi}{\d\b}\right)
\end{eqnarray}
or, equivalently, we arrange equations to obtain,
\begin{equation}\label{eqn F}
\frac{\d}{\d\a} \left(\frac{\d\chi}{\d\a}-\frac{Y_2(h u_\a)}{1-\s} \right) + \frac{\d}{\d\b} \left(\frac{\d\chi}{\d\b}-\frac{Y_2(h u_\b)}{1-\s}\right)=0,
\end{equation}
from $\s_{\a\a}+\s_{\b\b}$ and $\s_{\a\a}-\s_{\b\b}$, respectively.
There exists a function $G$ which satisfies equation (\ref{eqn F}) for which $\frac{\d G}{\d \b}= \frac{\d\chi}{\d\a}-\frac{Y_2}{1-\s}h u_\a$ and $\frac{\d G}{\d\a}=-\frac{\d \chi}{\d\b}+\frac{Y_2}{1-\s} h u_\b$.  Using these equations, we eliminate $u_\a$ and $u_\b$ from equation (\ref{saa-sbb})
\begin{equation}
\frac{\d^2}{\d\a\d\b}\left(\frac{G}{h}\right)=-\frac{h}{1+\s}\left(\s_{\a\a}-\s_{\b\b}\right),
\end{equation}
the left hand side of this equation may be written in this manner because $\frac{\d^2}{\d\a\d\b}\frac{1}{h}=0$.  Thus, the relative displacement vectors are given by
\begin{eqnarray}\label{bipolar displacements}
u_\a&=&\frac{1-\s}{Y_2}\frac{1}{h}\left(\frac{\d\chi}{\d\a}-\frac{\d G}{\d\b}\right)\nonumber\\
u_\b&=& \frac{1-\s}{Y_2}\frac{1}{h}\left(\frac{\d\chi}{\d\b}+ \frac{\d G}{\d\a}\right).
\end{eqnarray}

\subsection{Fixing a Hole  {${\textrm{\footnotesize  \textit{ (Outtake)}}}$}}
Recall, $\chi$ is the solution to the biharmonic equation $\nabla ^4 \chi=0$ with respect to the symmetries in our system.  The biharmonic equation when written in terms of the function $\chi/h$ has the simplified form,
\begin{equation}\label{biharmonic}
\left(\frac{\d^4}{\d \a^4}+ \frac{\d^4}{\d\b^4}+2 \frac{\d^4}{\d\a^2\d\b^2}-2 \frac{\d^2}{\d \a^2}+2 \frac{\d^2}{\d\b^2}+1 \right) \frac{\chi}{h}=0.
\end{equation}
Similarly, the components of the stress tensor are given by,
\begin{eqnarray}
\s_{\a\a} &=& \left[ \frac{1}{h}\frac{\d^2}{\d\b^2}-\frac{\sinh \a}{a} \frac{\d}{\d\a}-\frac{\sin \b}{a} \frac{\d}{\d\b}+\frac{\cosh\a}{a}\right]\frac{\chi}{h}\nonumber\\
\s_{\b\b}&=& \left[ \frac{1}{h}\frac{\d^2}{\d\a^2}-\frac{\sinh \a}{a} \frac{\d}{\d\a}-\frac{\sin \b}{a} \frac{\d}{\d\b}+\frac{\cos\b}{a}\right]\frac{\chi}{h}\nonumber\\
\s_{\a\b}&=&-\frac{1}{h} \frac{\d^2}{\d\a\d\b}\left(\frac{\chi}{h}\right).
\end{eqnarray}

Our system, while undergoing uniform hydrostatic compression, is described by the stress function in equation (\ref{airy bipolar 0}),
\begin{equation}
\frac{\chi^\0}{h}=\frac{P a}{2}\left(\cosh \a+ \cos \b\right),\nonumber
\end{equation}
or, equivalently, is given by the components of the stress tensor
\begin{equation}
\s_{\a\a}^\0=\s_{\b\b}^\0=P, \quad \s_{\a\b}^\0=0.
\end{equation}
We aim to find solutions to the biharmonic equation that are even in both $\a$ and $\b$.  Thus, the Airy stress function is given by
\begin{widetext}
\begin{eqnarray}
\frac{\chi^\1}{h}&=&C \left(\cosh \a-\cos\b\right) \log \left(\cosh \a-\cos\b\right)+ \sum_{n=1}^\infty \phi_n(\a)\cos n \b,
\end{eqnarray}
where $\phi_n(\a) = A_n \cosh(n+1)\a+B_n \cosh(n-1)\a$.  The components of the stress tensor corresponding to this Airy stress function are
\begin{eqnarray}\label{stress bipolar 1}
\s_{\a\a}^\1&=&-\frac{C}{2 a}\left(\cosh 2 \a-2\cosh \a \cos\b+\cos 2 \b\right) + \frac{1}{a}\phi_1(\a)\nonumber\\
&+&\frac{1}{2a}\sum_{n=1}^\infty \frac{1}{n}\left[ f_{n+1}(\a)-2 \cosh \a f_n(\a) +f_{n-1}(\a)- 2 \sinh \a g_n(\a)\right] \cos n\b \\
\s_{\b\b}^\1&=&\frac{C}{2 a} \left(\cosh 2 \a-2 \cosh \a \cos\b+\cos 2\b\right) + \frac{1}{a}\phi_1(\a)-\frac{1}{2 a}\phi_1''(\a)\nonumber\\
&-&\sum_{n=1}^\infty \left[\phi_{n+1}''(\a)-2 \cosh \a \phi_n''(\a)+ \phi_{n-1}''(\a)+(n+2) \phi_{n+1}(\a)+2 \sinh \a \phi_n'(\a)+(n-2)\phi_{n-1}(\a)\right] \cos n\b \quad \ \ \ \\
\s_{\a\b}^\1&=& -\frac{C}{a} \sinh \a \sin\b -\frac{1}{2 a} \sum_{n=1}^\infty \left[g_{n+1}(a)-2 \cosh \a g_n(\a)+g_{n-1}(\a) \right] \sin n\b,
\end{eqnarray}
\end{widetext}
where $f_n(\a)=(n+1)n(n-1)\phi_n(\a)$ and $g_n(\a)=n \phi_n'(\a)$.  The boundary conditions require there be no stress at infinity ($\a=0$), hence
\begin{equation}
\sum_{n=1}^\infty\left[A_n+B_n\right]=0,
\end{equation}
and the normal and tangential stresses must vanish along the edges of the holes, located at $\a=\pm s$.  Thus, the constants $A_n$, $B_n$ and $C$ must satisfy the following recurrence relations for $n\ge2$ 
\begin{eqnarray}
\label{fn} f_{n+1}(s)-2 \cosh s f_n(s)+f_{n-1}(s)&=&2 \sinh s g_n(s),\qquad \\
\label{gn} g_{n+1}(s)-2 \cosh s g_n(s)+g_{n-1}(s)&=&0,
\end{eqnarray}
subject to the conditions
\begin{eqnarray}
\label{cond1}2 \phi_1(s)&=&-2\frac{P}{a}-C \cosh 2 s,\\
\label{cond2}f_{2}(s)-2 \sinh s g_1(s)&=&2 \cosh s,\\s
\label{cond3}-2 \cosh s f_2(s)+f_3(s)&=&2 \sinh s g_2(s)+ 2 C,\qquad \ \ \\
\label{cond4}2 \cosh s g_1(s)-g_2(s)&=&2 C \sinh s.
\end{eqnarray}
Using equation (\ref{gn}), we find $g_n(s)=c_1 \lambda_1^n+c_2 \lambda_2^n$, where $\lambda_1$ and $\lambda_2$ are roots of the characteristic polynomial $t^{n+1}-2 \cosh s t^n+t^{n-1}=0,$ yielding $g_n(s)=c_1 e^{-n s}+c_2 e^{n s}$.  However, the stress must be finite everywhere, thus, $c_2=0$.  Equation (\ref{cond4}) completes the relation for $g_n(s)$ as $c_1=2 C \sinh s$.  Because, equation (\ref{fn}) is a non-linear recurrence relation, we must consider $f_{n+2}-2 \cosh s f_{n+1} s+f_n{s}-\frac{g_{n+1}}{g_{n}}\left(f_{n+1}(s)-2 \cosh s f_n(s)+f_{n-1}(s)\right)=0$.  Two of the roots of the characteristic polynomial for this equation are degenerate, the equation is $f_n(s)=d_1 e^{-ns}+d_2 n e^{-ns}+d_3e^{ns}$.  From the boundary conditions, we determine $f_n(s)=-2K\left( \cosh s+n \sinh s\right) e^{-ns}.$  From the definitions of $f_n(\a)$ and $g_n(\a)$, we find that the coefficients $A_n$, $B_n$ and $C$ satsify,
\begin{eqnarray}
A_n&=&2 C\frac{e^{- ns}\sinh n s+n e^{-s} \sinh s}{n(n+1)\left(\sinh 2 ns+n \sinh 2 s\right)},\nonumber\\
B_n&=&-2 C\frac{e^{- ns}\sinh n s+n e^{s} \sinh s}{n(n-1)\left(\sinh 2 ns+n \sinh 2 s\right)},\nonumber\\
& & \textrm{with} \ \  B_1=\frac{C}{2} \tanh s\cosh 2 s+P,\nonumber\\
\end{eqnarray}
and
\begin{widetext}
\begin{eqnarray}
C &=&- P\left\{\frac{1}{2}+\tanh s\sinh^2s -4 \sum_{n=2}^\infty  \left[\frac{e^{-ns}\sinh ns+n \sinh s \left(n \sinh s+\cosh s \right)}{n(n^2-1)\left( \sinh 2n s+n \sinh 2 s \right)}  \right]\right\}^{-1}.
\end{eqnarray}

Now that we have equations for the stresses everywhere, we may now solve for the field of relative displacement vectors.  Recall, our function $G$ is given by,
\begin{eqnarray}
G&=&\frac{h}{1+\s}\int \int d\a d\b \left\{\frac{\d^2}{\d\a^2}-\frac{\d^2}{\d\b^2}-1\right\}\frac{\chi}{h}\nonumber\\
&=&\frac{2h}{1+\s}\left[2 C\left(\tan^{-1}\left(\tanh\frac{\a}{2}\cot\frac{\b}{2}\right)\cos \b+\tan^{-1}\left(\coth\frac{\a}{2}\tan\frac{\b}{2}\right) \cosh \a \right)+\sum_{n=1}^\infty\psi_n(\a)\sin n\b \right],
\end{eqnarray}
where $\chi=\chi^\0+\chi^\1$ and $\psi(\a)=A_n \sinh(n+1)\a+B_n \sinh(n-1)\a$.  To calculate the displacement field, we will need to know the following relations,
\begin{eqnarray}
\frac{\d \chi}{\d\a}&=&h\left(\sinh \a( -P h \cos \b+C)-\sum_{n=1}^\infty \left[\frac{h}{a} \phi_n(\a)\sinh\a-\phi_n'(\a)\right] \cos n \b\right)\\
\frac{\d\chi}{\d\b}&=&h \left(\sin \b(-P h \cosh \a+C)-\sum_{n=1}^\infty\phi_n(\a)\left[\frac{h}{a}\sin \b \cos n \b+n \sin n \b\right]\right)\\
\frac{\d G}{\d\a}&=&-\frac{2 h}{1+\s} \left(C \left(\pi \frac{h}{a} \sinh \a \cos\b +\sin \b\right)+\sum_{n=1}^\infty \left[\frac{h}{a}\psi_n(\a) \sinh \a -\psi_n'(\a)\right] \sin n\b \right)\\
\frac{\d G}{\d\b}&=&-\frac{2h}{1+\s}\left(C \left(\pi \frac{h}{a}\cosh \a \sin \b-\sinh \a \right)  +\sum_{n=1}^\infty \psi_n(\a)\left[\frac{h}{a}\sin \b \sin n \b-n \cos n \b\right] \right).
\end{eqnarray}
\end{widetext}
These equations, together with Eqs. (\ref{bipolar displacements}), complete our description of the system of two holes under hydrostatic compression, which may be seen in Fig. \ref{two holes}.  It should be noted that in the compressed system there is an instability for large enough values of $P$ wherein the displacements intersect each other, causing overlap in the system.

Due to the nature of problems in the theory of linear elasticity theory, the most useful identities involve the directional cosines relating the $\{\a,\b\}$ to the $\{x,y\}$ coordinates.  The unit vectors in the new system are given by $\hat{\boldsymbol{\a}} =\frac{1}{h_\a} \big(\frac{\partial x}{\partial \a} \hat{\boldsymbol{x}} +\frac{\partial y}{\partial \a} \bf{\hat{\boldsymbol{y}}}\big)$ and $\hat{\boldsymbol{\b}} =\frac{1}{h_\b} \big(\frac{\partial x}{\partial \b} \hat{\boldsymbol{x}} +\frac{\partial y}{\partial \b} \hat{\boldsymbol{y}}\big)$.  The orthogonality condition $\hat{\boldsymbol{\a}} \cdot \hat{\boldsymbol{\b}} =0$ implies $\frac{\partial x}{\partial \a} \frac{\partial x}{\partial \b}+\frac{\partial y}{\partial \a} \frac{\partial y}{\partial \b}=0.$  The directional cosines are related because $h_\a ^2= \big(\frac{\d x}{\d \a} \big)^2 \big(1+\big(\frac{\d y}{\d \a}/ \frac{\d x}{\d \a} \big)^2\big) =\big(\frac{\d x}{\d \a} \big)^2 \big(1+\big(-\frac{\d x}{\d \b}/ \frac{\d y}{\d \b} \big)^2\big) = h_\b^2 \big(\frac{\d x}{\d \a} \big)^2\big(\frac{\d y}{\d \b} \big)^{-2},$ or
\begin{eqnarray}\label{id1}
\frac{1}{h_\a} \frac{\d x}{\d \a} = \frac{1}{h_\b} \frac{\d y}{\d \b}, \quad  \frac{1}{h_\a} \frac{\d y}{\d \a} = -\frac{1}{h_\b} \frac{\d x}{\d \b},
\end{eqnarray}
where the sign is chosen such that both $\{\hat{\boldsymbol{x}},\hat{\boldsymbol{y}},\hat{\boldsymbol{z}}\}$ and $\{\hat{\boldsymbol{\a}},\hat{\boldsymbol{\b}},\hat{\boldsymbol{z}}\}$ form right-handed orthonormal triads.
Thus, under the change of coordinates $\{x,y,z\}\rightarrow \{\a,\b,z\}$, a rank-2 tensor $A_{ij} = a_{ij}\hat{\boldsymbol{x}_i} \hat{\boldsymbol{x}_j}$ transforms as $A_{\mu \nu} = a_{\mu \nu} \hat{\boldsymbol{\xi}_\mu} \hat{\boldsymbol{\xi}_\mu}= a_{ij} \left(\hat{\boldsymbol{\xi}_\mu} \cdot \hat{\boldsymbol{x}_i} \right) \hat{\boldsymbol{\xi}_\mu}  \left(  \hat{\boldsymbol{\xi}_\nu} \cdot \hat{\boldsymbol{x}_j} \right) \hat{\boldsymbol{\xi}_\nu}$, which may be written as
\begin{equation}
a_{\mu \nu} = \frac{1}{h_\mu h_\nu} \frac{\d x_i}{\d \xi_\mu} \frac{\d x_j}{\d \xi_\nu} a_{ij}.
\end{equation}

The derivatives of the directional cosines can be made from linear combinations of derivatives of the orthogonality condition $\frac{\partial x}{\partial \a} \frac{\partial x}{\partial \b}+\frac{\partial y}{\partial \a} \frac{\partial y}{\partial \b}=0$ and the scale functions.  For example, to find $\frac{\d}{\d \a} \big( \frac{1}{h_\a} \frac{\d x}{\d \a} \big),$ first note there are two ways of obtaining this derivative; directly,
\begin{equation}\label{deriv1}
\frac{\d}{\d \a} \left( \frac{1}{h_\a} \frac{\d x}{\d \a} \right) = -\frac{1}{h_\a^2} \frac{\d h_\a}{\d \a} \frac{\d x}{\d \a} + \frac{1}{h_\a}\frac{\d^2 x}{\d \a^2},
\end{equation}
and by taking the derivative of the product of $\a$ scale function and orthogonality condition with respect to $\a$
\begin{widetext}
\begin{eqnarray}\label{deriv2}
\frac{\d}{\d \a} \left( \frac{1}{h_\a} \frac{\d x}{\d \a} \right) \frac{\d x}{\d b} &=&\frac{1}{h_\a}\left( \frac{1}{h_\a}\frac{\d h_\a}{\d \a} \frac{\d y}{\d \a}\frac{\d y}{\d \b} - \frac{\d^2 y}{\d \a^2}\frac{\d y}{\d \b} -\frac{\d^2 x}{\d \a \d \b}\frac{\d x}{\d \a} -\frac{\d^2 y}{\d \a \d \b}\frac{\d y}{\d \a} \right)\nonumber\\
&=&  \frac{1}{h_\a^2}\frac{\d h_\a}{\d \a} \frac{\d y}{\d \a}\frac{\d y}{\d \b} - \frac{1}{h_\a} \frac{\d^2 y}{\d \a^2}\frac{\d y}{\d \b} - \frac{\d h_\a}{\d \b},
\end{eqnarray}
where we have made use of the definition $\frac{\d h_\a}{\d \b} = \frac{\d^2 x}{\d \a \d \b}\frac{\d x}{\d \a}+\frac{\d^2 y}{\d \a \d \b}\frac{\d y}{\d \a}.$  Next multiplying Eqn. \ref{deriv1} by $\frac{\partial x}{\d \b}$ and Eqn. \ref{deriv2} by $ \big( \frac{\d y}{\d \b}\big)^2$ and taking their sum, this becomes
\begin{eqnarray}
h_\b^2 \frac{\d}{\d \a} \left( \frac{1}{h_\a} \frac{\d x}{\d \a} \right)  &=&\frac{1}{h_\a^2}\frac{\d h_\a}{\d \a} \frac{\d y}{\d \b} \left( \frac{\d y}{\d \a}\frac{\d x}{\d \b} - \frac{\d y}{\d \b}\frac{\d x}{\d \a} \right) - \frac{\d h_\a}{\d \b} \frac{\d x}{\d \b} + \frac{1}{h_\a^2} \frac{\d y}{\d \b} \left( \frac{\d y}{\d \b}\frac{\d^2 x}{\d \a^2} - \frac{\d x}{\d \b}\frac{\d^2 y}{\d \a^2} \right) \nonumber\\
 &=& \frac{1}{h_\a^2} \frac{\d h_\a}{\d \a}\frac{\d y}{\d \b}\frac{h_\b}{h_\a} - \frac{\d h_\a}{\d \b}\frac{\d x}{\d \b} - \frac{1}{h_\a}\frac{\d y}{\d \b} \frac{h_\b}{h_\a} \frac{\d h_\a}{\d \a}.
\end{eqnarray}
By following a the same procedure, the formul\ae\ for the derivatives of the directional cosines are
\begin{eqnarray}\label{id3}
\frac{\d}{\d \a} \left( \frac{1}{h_\a}\frac{\d x_i}{\d \a} \right) &=& - \frac{1}{h_\b^2} \frac{\d h_\a}{\d \b} \frac{\d x_i}{\d \b}, \quad
\frac{\d}{\d \b} \left( \frac{1}{h_\a}\frac{\d x_i}{\d \a} \right) = \frac{1}{h_\a h_\b} \frac{\d h_\b}{\d \a} \frac{\d x_i}{\d \b}, \nonumber\\
\frac{\d}{\d \a} \left( \frac{1}{h_\b}\frac{\d x_i}{\d \b} \right) &=& \frac{1}{h_\a h_\b} \frac{\d h_\a}{\d \b} \frac{\d x_i}{\d \a}, \quad
\frac{\d}{\d \b} \left( \frac{1}{h_\b}\frac{\d x_i}{\d \b} \right) = - \frac{1}{h_\a^2} \frac{\d h_\b}{\d \a} \frac{\d x_i}{\d \a}, 
\end{eqnarray}
\end{widetext}
where $i=1,2$ and $x_1=x$ and $x_2=y$.  Note that the cross partial derivatives of the directional cosines are equal,
\begin{eqnarray}
\frac{\d}{\d \b} \left(\frac{\d}{\d \a} \left( \frac{1}{h_\a}\frac{\d x_i}{\d \a} \right) \right) - \frac{\d}{\d \a} \left( \frac{\d}{\d \b} \left( \frac{1}{h_\a}\frac{\d x_i}{\d \a} \right) \right)&=&0,\quad \ \ \ \nonumber\\
\frac{\d}{\d \b} \left(\frac{\d}{\d \a} \left( \frac{1}{h_\b}\frac{\d x_i}{\d \b} \right) \right) - \frac{\d}{\d \a} \left( \frac{\d}{\d \b} \left( \frac{1}{h_\b}\frac{\d x_i}{\d \b} \right) \right)&=&0,\quad \ \ \
\end{eqnarray}
which leads to our final identity
\begin{equation}
\frac{\d}{\d \b}\left(\frac{1}{h_\b}\frac{\d h_\a}{\d \b} \right) + \frac{\d}{\d \a}\left( \frac{1}{h_\a}\frac{\d h_\b}{\d\a}\right)=0.
\end{equation}

\subsection{Identities in Orthogonal Coordinates}

We include these for completeness.  These were necessary for us to study the Airy stress formalism in orthogonal coordinates and we did not find these, presumably known identities, in any reference.

Consider the general set of orthogonal coordinates, $\{\a(x,y),\b(x,y)\}$.  The new basis preserves length of the differential line element, $ds^2 = dx^2 + dy^2 = h_\a^2 d\a^2 + h_\b^2  d\b^2= h_\a^2 \big(\frac{\partial \a}{\partial x} dx + \frac{\partial \a}{\partial y} dy \big)^2 +h_\b^2 \big(\frac{\partial \b}{\partial x} dx + \frac{\partial \b}{\partial y} dy \big)^2,$ defining the scale functions $h_\a^{-2} =  \big( \frac{\partial \a}{\partial x} \big)^2 + \big( \frac{\partial \a}{\partial y} \big)^2$ and $h_\b^{-2} =  \big( \frac{\partial \b}{\partial x} \big)^2 + \big( \frac{\partial \b}{\partial y} \big)^2.$  However, it is often more useful to consider Cartesian coordinates as functions of the new orthogonal ones, $\{x(\a,\b),y(\a,\b)\}$, which yield an equivalent statement of the scale functions $h_\a^2=\big( \frac{\partial x}{\partial \a} \big)^2 + \big( \frac{\partial y}{\partial \a} \big)^2$ and $h_\b^2=\big( \frac{\partial x}{\partial \b} \big)^2 + \big( \frac{\partial y}{\partial \b} \big)^2.$  By transforming from orthogonal back to Cartesian coordinates, the differential line element gives $h_\a^2 \big( \frac{\partial x}{\partial \a} \big)^2 + h_\b^2 \big( \frac{\partial x}{\partial \b} \big)^2 =1$ and $h_\a^2 \big( \frac{\partial y}{\partial \a} \big)^2 + h_\b^2 \big( \frac{\partial y}{\partial \b} \big)^2 =1$; from which, we obtain the first set of identities:
\begin{eqnarray}\label{id2}
\frac{\partial x}{\partial \a}&=&h_\a^2 \frac{\partial \a}{\partial x}, \quad \frac{\partial y}{\partial \a}=h_\a^2 \frac{\partial \a}{\partial y}\nonumber\\
\frac{\partial x}{\partial \b}&=&h_\b^2 \frac{\partial \b}{\partial x}, \quad \frac{\partial y}{\partial \b}=h_\a^2 \frac{\partial \b}{\partial y}.
\end{eqnarray}

\section{Filling the Cracks: The Dislocation Dipole}
We devote this Appendix to the technical details of the mathematical manipulation required to compute the energetics of a lattice of dislocation dipoles. 

 \subsection{The Dipole Term}\label{dipole}
Despite the complex form of the integral in the dislocation dipole interaction energy, when completed it has a surprisingly simple form.  The integral becomes tractable by transforming to polar coordinates, and then manipulating the trigonometric functions.  In polar coordinates, the integral in Eq. (\ref{burgers integral}) becomes,
\begin{widetext}
\begin{equation}
E=\frac{Y_2 b^2 d_1 d_2}{(2\pi)^2}\int_0^{2 \pi} d\theta \int_0^{\infty} q \, d q \, \cos \big(-q \, R \, \cos (\theta) \big) \cos^2(\theta-\theta_1) \cos^2(\theta-\theta_2)
\end{equation}
\end{widetext}
Employing Bessel function and trigonometric identities, we find
\begin{equation}
E = - \frac{Y_2}{\pi} \frac{b^2 \, d_1 \, d_2}{R^2} \left( \cos(\th_1+\th_2)\sin \th_1 \sin \th_2+\frac{1}{4}\right).
\end{equation}

\subsection{Why We Can Ignore Higher Order Terms}\label{higher order}

\begin{figure}[!hbt]
\begin{center}
\includegraphics[width=3.5in]{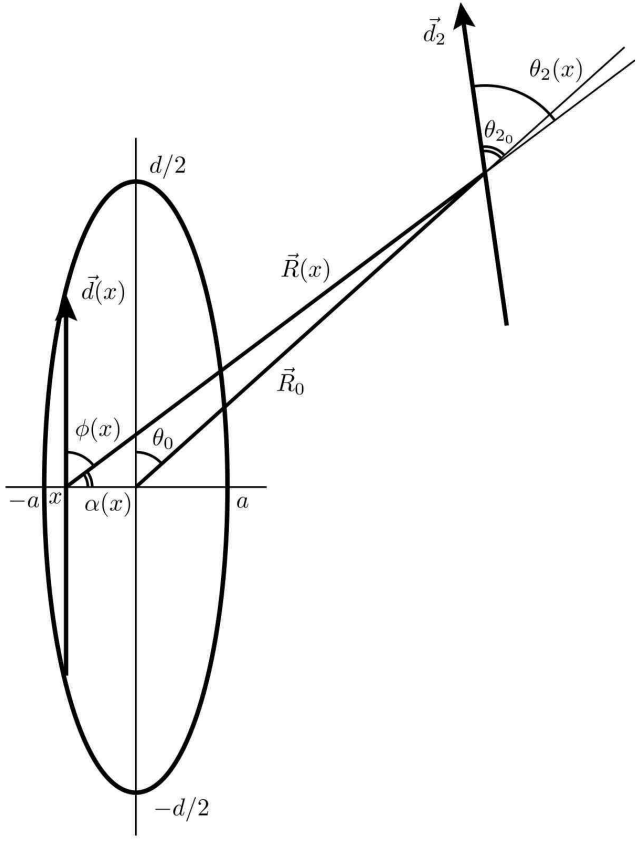}
\caption{Set up for dipole expansion of an elongated shape symmetric about both the $x$- and $y$-axes, in this case, an ellipse.}
\label{dipole expansion}
\end{center}
\end{figure}

Our goal is to prove that we need only consider the first order dipole-dipole term when considering collapsed holes.  For simplicity sake, we will study an elongated shape that is symmetric about both its semi-major and semi-minor axes, (see FIG. \ref{dipole expansion}).  Here, we consider only shapes for which the ratio of minor and major axes $2 a/d \ll 1.$  In the theory of cracks, the height profile of a crack $h(x)$ may be constructed from a continuous distribution of finite parallel edge dislocations.  A finite edge dislocation of length $\ell$ can be thought of as a two infinite edge dislocations of the same strength but opposite charge which is given by Burgers vectors situated at $\boldsymbol{b}_{+} = b \, \delta(\boldsymbol{x}-\boldsymbol{\ell}/2)$ and $\boldsymbol{b}_{-} = - b \, \delta(\boldsymbol{x}+\boldsymbol{\ell}/2),$ in other words, a dislocation dipole.  Given the height profile of a shape, it is trivial to construct it from such dislocation dipoles, $h(x) = d(x).$  Thus, the interaction energy between such a shape made from dislocation dipoles and a single dislocation dipole $\boldsymbol{d}_2$ located a distance $\boldsymbol{R}_0$ away is given by:
\begin{widetext}
\begin{equation}
E = - \frac{Y_2 \, b^2 \, d_2}{\pi}  \int_{-a}^{a} dx \, \frac{d(x)}{R^2(x)} \left( \cos \left(\f(x)+\th_2(x) \right) \sin \left( \f(x) \right) \sin \left( \th_2(x) \right) + \frac{1}{4} \right).
\end{equation}
By repeated application of the law of cosines, the functional forms of $R(x),$ $\f(x)=\pi/2-\alpha(x),$ and $\th_2(x)$ are:
\begin{eqnarray}
R^2(x) &=& R^2_0+x^2-2\,  R_0 \, x \cos \left( \frac{\pi}{2}+\th_0\right) = R^2_0+x^2+2\,  R_0 \, x \sin \th_0\\
\alpha(x) &=& \cos^{-1} \left( \frac{x + R_0 \sin \th_0}{\sqrt{R_0^2+x^2+2\, R_0 \, x \sin \th_0 }}\right)\\
\th_2(x) &=& \th_{2_0} -\th_0-\alpha(x)+\pi/2.
\end{eqnarray}
The energy density is 
\begin{eqnarray}
f =-\frac{Y_2 \, b^2 \, d_2}{\pi} \frac{ d(x)}{R^2_0+x^2+2\,  R_0 \, x \sin \th_0 } \left(- \cos (\tilde{\th} - 2 \, \alpha(x) ) \cos( \alpha(x)) \cos(\tilde{\th}-\alpha(x) )+ \frac{1}{4} \right),
\end{eqnarray}
where $\tilde{\th} = \th_{2_0}-\th_0.$
This may be vastly simplified by expanding the angular terms:
\begin{eqnarray}
\cos (\tilde{\th} - 2 \, \alpha(x) ) &=& \frac{\cos \tilde{\th} (x+R_0 \sin \th_0)^2 R_0^2 \cos^2 \th_0+2 \, \sin \tilde{\th}\left(x+R_0 \sin \th_0\right)R_0 \cos \th_0}{R_0^2+x^2+2 \, R_0 \, x \sin \th_0}\\
\cos (\tilde{\th} - \alpha(x) ) &=& \frac{ \cos \tilde{\th} \left(x+R_0 \sin \th_0\right)+ \sin \tilde{\th} R_0 \cos \th_0}{\sqrt{R_0^2+x^2+2 \, R_0 \, x \sin \th_0}}.
\end{eqnarray}
The energy density becomes:
\begin{eqnarray}
f &=& -\frac{Y_2 \, b^2 \, d_2}{\pi}\frac{ d(x)}{\left(R^2_0+x^2+2\,  R_0 \, x \sin \th_0\right)^3}\left(\frac{\left(R^2_0+x^2+2\,  R_0 \, x \sin \th_0\right)^2}{4}-\sum_{n=1}^{4} c_n (x+A)^n \right)\nonumber\\
&=& -\frac{Y_2 \, b^2 \, d_2}{\pi}\frac{ d(x)}{\left(R^2_0+x^2+2\,  R_0 \, x \sin \th_0\right)^3}\sum_{n=0}^{4} c_n \sum_{m=0}^n\frac{n!}{m!(n-m)!}A^{n-m} x^m 
\end{eqnarray}
where $A=R_0 \sin \th_0$ and $B= R_0 \cos \th_0,$ and the coefficients $c_n$ are given by $c_0=B^4/4,$ $c_1=B^3 \, \cos \tilde{\th} \, \sin \tilde{\th},$ $c_2=B^2( \cos^2 \tilde{\th}-2 \, \sin^2 \tilde{\th})+B^2/2,$ $c_3=- 3 \, B \, \cos \tilde{\th} \, \sin \tilde{\th},$ and $c_4=- \cos^2 \tilde{\th}+1/4.$
While the above energy is for general shape of dislocations, we choose a shape to do the actual calculation.  For simplicity sake, we choose an ellipse of major axis $d_0$ and minor axis $2 a$.  Thus, $d(x) = d_0 \sqrt{1-x^2/a^2}$.  With the change of variables, $y=x/a,$ our energy integral becomes:
\begin{equation}
E= - \frac{Y_2 \, b^2 \, d_2}{\pi\, R_0^6} \int_{-1}^{1}dy\frac{d_0 \, a \sqrt{1-y^2}}{\left(1+(\frac{a}{R_0}  y)^2 + 2  \frac{a}{R_0}  y \, \sin \th_0 \right)^3} \sum_{n=0}^4c_n \sum_{m=0}^n \frac{n!}{m!(n-m)!}A^{n-m}(a \, y)^m.
\end{equation}
Expanding the denominator for $\frac{a}{R_0} \ll 1,$ we find, $\left(1 +(\frac{a}{R_0} \, y)^2 + 2 \frac{a}{R_0} y \, \sin \th_0 \right)^{-3} \approx \left(1- 6 \frac{a}{R_0} \sin \th_0 \, y+ 3 \big( \frac{a}{R_0}\big)^2 (8 \, \sin^2 \th_0-1) y^2 + \cdots \right).$  Now we need only do the integral $\int_{-1}^1 dy \sqrt{1-y^2}\,  y^n= 2(1-(-1)^n) \int_0^1 \sqrt{1-y^2} y^{N},$ for, since $\sqrt{1-y^2}$ is even, this integral is zero for odd integer $n$  This is very simple using beta functions, which we may see by the change of variables $t=y'^2$:
\begin{eqnarray}
2 \int_0^{1} dy' \, \sqrt{1-y'^2} \, y'^{N}  &=& \int_0^{1} \frac{dt}{\sqrt{t}} \sqrt{1-t} \, t^{n/2} = \int_0^{1} dt \, t^{(n+1)/2-1} (1-t)^{3/2-1}\nonumber\\
 &\equiv &B\Big(\frac{1+n}{2},\frac{3}{2}\Big) = \frac{\Gamma\left(\frac{1+n}{2}\right) \, \Gamma\left(\frac{3}{2} \right)}{\Gamma\left(2+\frac{n}{2}\right)} = \frac{\pi(1+(-1)^n)}{2^{\frac{n}{2}+1}} \frac{(n -1)!!}{(\frac{n}{2}+1)!}
 \end{eqnarray}
The interaction energy is thus:
\begin{eqnarray}
E&=& - \frac{Y_2 \, b^2 \, d_2\, d_0 \, a}{\pi\, R_0^6} \sum_{n=0}^4 \sum_{m=0}^n c_n \frac{n!}{m!(n-m)!} A^{n-m} a^m\frac{\pi }{2^{m/2+2}} \left((1+(-1)^m) \left( 2 \frac{(m-1)!!}{(\frac{m}{2}+1)!}\right. \right. \nonumber\\ 
&+&\left. \left. 3 \frac{a^2}{R_0^2} \left(8 \sin^2 \th_0-1 \right) \frac{(m+1)!!}{(\frac{m}{2}+2)!} \right)
- 6 \frac{a}{R_0} \sin \th_0 \sqrt{2} (1-(-1)^{m}) \frac{m!!}{(\frac{m+1}{2}+1)!} \right).
\end{eqnarray}
While this appears to be a complicated expression, let us, for the moment consider only the first order term in $a$:
\begin{eqnarray}
E_0&=&- \frac{Y_2 \, b^2 \, d_2\, d_0 \, a}{4 \, R_0^6} \sum_{n=0}^4 c_n \, A^{n} = - \frac{Y_2 \, b^2 \, d_2\, d_0 \, a}{4 \, R_0^6}  \left( \frac{B^4}{4}+A \, B^3 \cos \tilde{\th} \sin \tilde{\th}\right. \nonumber\\
&+&\left. A^2 \, B^2 \left( \cos^2 \tilde{\th}-2 \sin^2 \tilde{\th} +\frac{1}{2} \right) - 3 \, A^3 \, B \cos\tilde{\th} \sin\tilde{\th}+ A^4 \left(\frac{1}{4}- \cos ^2\tilde{\th}  \right) \right)\nonumber\\
&=&- \frac{Y_2 \, b^2 \, d_2\, d_0 \, a}{4 \, R_0^2} \left(\frac{\cos^4\th_0}{4}  + \left(\cos^3 \th_0 \sin \th_0 -3 \sin^3 \th_0 \cos \th_0 \right)\cos \tilde{\th} \sin \tilde{\th} \right. \nonumber\\
&+&\left. \cos^2 \th_0 \sin^2 \th_0 \left( \cos^2 \tilde{\th}- 2 \sin^2 \tilde{\th} +\frac{1}{2} \right) + \sin^4  \th_0 \left( \frac{1}{4}-\cos^2 \tilde{\th} \right)\right).
\end{eqnarray}
This does not appear to have the same functional form as the original energy.  However, with the help of some trigonometric identities, we begin to simplify the energy:
\begin{eqnarray}
& & \cos^4 \th_0+ \sin^4 \th_0 \left( 1-4 \cos^2\tilde{\th} \right)= \frac{1}{2}\sin^2 2 \th_0 \cos 2 \tilde{\th}+1 - 4 \sin^2 \th_0 \cos^2 \tilde{\th}\\
& &  \big(\cos^3 \th_0 \sin \th_0 -3 \sin^3 \th_0 \cos \th_0 \big)\cos \tilde{\th} \sin \tilde{\th} = \frac{1}{4} (2\, \cos 2\th_0-1) \sin 2 \th_0 \sin 2 \tilde{\th}\\
& &\Big( \cos^2 \tilde{\th}- 2 \sin^2 \tilde{\th} +\frac{1}{2} \Big)  \cos^2 \th_0 \sin^2 \th_0 = \frac{3}{8} \cos 2 \tilde{\th}  \sin^2 2 \th_0.
\end{eqnarray}
We now use these identities in the energy to find:
\begin{eqnarray}
E_0&=&- \frac{Y_2 \, b^2 \, d_2\, d_0 \, a}{4 \, R_0^2} \left(\frac{1}{8}\sin^2 2 \th_0 \cos 2 \tilde{\th}+\frac{1}{4} - \sin^2 \th_0 \cos^2 \tilde{\th}+\frac{1}{4} (2\, \cos 2\th_0-1) \sin 2 \th_0 \sin 2 \tilde{\th} + \frac{3}{8} \cos 2 \tilde{\th}  \sin^2 2 \th_0 \right)\nonumber\\
 &=& - \frac{Y_2 \, b^2 \, d_2\, d_0 \, a}{4 \, R_0^2} \left( \frac{1}{2} \sin 2 \th_0 \left(\sin 2\th_0 \cos 2 \tilde{\th} + \cos 2 \th_0 \sin 2 \tilde{\th} \right)- \sin^2 \th_0 \cos^2 \tilde{\th} -\frac{1}{4} \sin 2 \th_0 \sin 2 \tilde{\th} + \frac{1}{4} \right).
\end{eqnarray}
Now we substitute $\tilde{\th} = \th_{2_0}-\th_0$ to find,
\begin{eqnarray}
E_0&=& - \frac{Y_2 \, b^2 \, d_2\, d_0 \, a}{4 \, R_0^2} \left(  \frac{1}{2} \sin 2 \th_0 \sin2 \th_{2_0} - \sin^2 \th_0 \cos^2 \tilde{\th} - \sin \th_0 \cos \th_0 \sin \tilde{\th} \cos \tilde{\th} + \frac{1}{4} \right)\nonumber\\
&=&- \frac{Y_2 \, b^2 \, d_2\, d_0 \, a}{4 \, R_0^2} \left(  \cos( \th_0 + \th_{2_0} ) \sin \th_0 \sin \th_{2_0}+ \frac{1}{4} \right).
\end{eqnarray}
\end{widetext}
So we have found that to first order in $a/R_0$ an ellipse of major axis $d$ and minor axis $2 a$ interacts with a dislocation dipole of strength $d_2$ a distance $R_0$ away like a dislocation dipole with effective dipole strength $\tilde{d} =  \pi \, d_0\, a /4.$  Higher order terms contain more powers of $a/R_0$, which can be neglected for very thin shapes or for dipoles that are very far away.


\begin{thebibliography}{99}
\bibitem{note1}  We use the convention $x_i$ to represent the $i^{\rm{th}}$ component of a Cartesian coordinate system, where the subscript can be any Roman letter.  Components of another orthogonal coordinate system are denoted by $\xi_\mu$, for any Greek subscript.
\end{thebibliography}
\end{document}